\documentclass[aps,preprint,amsmath,amssymb]{revtex4}
\usepackage{graphicx}
\def\gq{g^{Q}}
\def\gl{g^{L}}
\def\g5{\gamma_{5}}
\def\ga{\gamma}
\def\ve{\varepsilon}
\def\la{\lambda}
\def\qp{q^{\prime}}

\begin{document}
\title{\Large \bf Implications of the HyperCP Data on $B$ and $\tau$ decays}
\date{\today}

\author{ \bf \large Chuan-Hung Chen$^{1,2}$\footnote{Email:
physchen@mail.ncku.edu.tw} and Chao-Qiang
Geng$^{3,4}$\footnote{Email: geng@phys.nthu.edu.tw}
 }

\affiliation{ $^{1}$Department of Physics, National Cheng-Kung
University, Tainan 701, Taiwan \\
$^{2}$National Center for Theoretical Sciences, Taiwan\\
$^{3}$Department of Physics, National Tsing-Hua University, Hsinchu
300, Taiwan  \\
$^{4}$Theory Group, TRIUMF, 4004 Wesbrook Mall, Vancouver, B.C V6T 2A3, Canada
 }

\begin{abstract}
If the HyperCP three events for the decay of $\Sigma^+\to
p\mu^+\mu^-$ are explained by a new pseudoscalar (axial-vector)
boson $X_{P(A)}$ with a mass of $214.3$ MeV, we study the
constraints on the couplings between $X_{P(A)}$ and fermions from
the experimental data in $K$ and $B$ processes. Some implications of
the new particle on  flavor changing  $B$ and $\tau$ decays are
given. Explicitly, we show that the decay branching ratios of
$B_{s}\to\phi X_{P(A)}\to \phi\mu^{+}\mu^{-}$, $B_{d}\to
K_{0}^{*}(1430){P(A)}\to K_{0}^{*}(1430)\mu^{+}\mu^{-}$ and
$\tau\to\mu X_{P(A)}\to\mu\mu^{+}\mu^{-}$ can be as large as $2.7\
(2.8)\times 10^{-6}$, $7.4\ (7.5)\times 10^{-7}$ and $1.7\
(0.14)\times 10^{-7}$, respectively.

\end{abstract}
\maketitle

The HyperCP collaboration \cite{Park}
has presented the branching ratio of
$\Sigma^+\to p\mu^+\mu^-$ to be $(8.6^{+6.6}_{-5.4}\pm 5.5)\times
10^{-8}$,
which is hardly explained within
the Standard Model \cite{Park,Bergstrom,He},
 and suggested a new boson $X$ with a mass of $214.3\pm
0.5$ MeV to induce the flavor changing transition of $s\to
d\mu^+\mu^-$.
It has been demonstrated \cite{HTV_PLB,DEJ,GH}
that to explain the data
 the new particle cannot be a scalar (vector)
 but pseudoscalar (axial-vector) boson $X_{P(A)}$ based on the direct
constraints from $K^+\to\pi^+\mu^+\mu^-$ and $K_L\to\mu^+\mu^-$. A
possible candidate with a light sgoldstino in spantaneously local
supersymmetry breaking theories has been extensively discussed in
the literature \cite{GR}. Recently, He, Tandean and Valencia
\cite{HTV-LH} have also shown that the light pseudoscalar Higgs
boson in the next-to-minimal supersymmetric standard model can be
identified as $X_{P}$. In addition, searching for the new light
boson at colliders has been studied by Ref.~\cite{Zhu}.

  In this paper, we will explore
  the implications of the HyperCP Data on
 flavor changing  $B$ and $\tau$ decays. In particular, we will
 examine constraints on the effective interactions induced by $X_{P,A}$
 from the experimental data in $B$ processes and study the possibility
 of having large effects in semileptonic $B_{d,s}$ decays. For the tau decays, we will concentrate on the decays of
 $\tau\to \ell \mu^{+}\mu^{-}$.

We start by writing the effective interactions for the new
pseudoscalar $(X_{P})$
or axial-vector $(X_{A})$ particle coupling to quarks and leptons to be
\cite{HTV_PLB}
\begin{eqnarray}
-{\cal L}_{P}=\left(-i\gq_{Pij}  \bar{q}_{i} \g5 q_{j} -i \gl_{Pij}
\bar{\ell}_{i} \g5 \ell_{j} +H.c. \right)X_{P}\,, \nonumber \\
-{\cal L}_{A}=\left(\gq_{Aij}  \bar{q}_{i} \ga_{\mu}\g5 q_{j} +
\gl_{Aij} \bar{\ell}_{i} \ga_{\mu}\g5 \ell_{j} +H.c.
\right)X^{\mu}_{A}\,, \label{eq:effint}
\end{eqnarray}
where
$g^{F}_{Pij}(g^{F}_{Aij})\ (F=Q,L)$ denote the
couplings of $X_{P}(X^{\mu}_A)$ to quarks and leptons, respectively,
and the
 indices $i,j$ stand for the quark or lepton flavors.
Although the exotic events observed in the HyperCP are associated with
the flavor changing neutral current (FCNC)
 in the first two generations of quark flavors and lepton flavor
(LF) conservation, to study the effects of the new particle on $B$
and $\tau$ decays, we will include all FCNCs in both quark and
lepton sectors. It has been studied and known that the constraint
 on the $s-d-X$ coupling from the decay branching ratio (BR) of $K_L\to \mu^{+} \mu^{-}$
 is more
strict than that from the $K-\bar K$ mixing \cite{HTV_PLB}. In order to
search for the most strict bounds on X-mediated $B$-meson decay
processes, we will examine those
measured
well in experiments, such as the $B_{d(s)}-\bar B_{d(s)}$ mixings and the decays of
$B_{d,s}\to \mu^{+} \mu^{-}$ and  $B\to K^{*} \mu^{+} \mu^{-}$.

To study the $B_q$ related processes, in terms of
Eq.~(\ref{eq:effint}), we explicitly write the relevant interactions
to be
\begin{eqnarray}
-{\cal L}=-i\gq_{Pbq}\bar b \g5 q X_{P}-i\gl_{P\mu}\bar \mu\g5 \mu X_{P}+
\gq_{Abq} \bar b \ga_{\mu} \g5 q X^{\mu}_{A}+\gl_{A\mu}
\bar\mu\ga_{\mu} \g5 \mu X^{\mu}_{A}+H.c.\,. \label{eq:intbq}
\end{eqnarray}
We note that for simplicity, we have abbreviated  $\gq_{Pbq}$
($\gq_{Abq}$) to denote the effective coupling for $b-q-X_{P(A)}$.
>From Eq. (\ref{eq:intbq}),
 the effective Hamiltonian for $|\Delta B|=2$
could be obtained by
\begin{eqnarray}
H^{|\Delta B|=2}_{\rm
eff}=\frac{g^2_{X}}{m^2_{B_q}-m^2_{X}}\left(\bar b \Gamma \g5 q
\right)\left(\bar b \Gamma \g5 q \right)\,,
\end{eqnarray}
where $g_{X}=(\gq_{Pbq},\, \gq_{Abq})$, $m_{X}=(m_{X_P},\, m_{X_A})$
and $\Gamma=(1,\, \ga_{\mu})$ for $X=(X_P,\, X^{\mu}_{A})$.
 The $B_q-\bar B_q$ oscillations are dictated by
the two physical mass differences, defined by $\Delta
m_{B_q}=2|M^{\Delta B=2}_{12}|=2|\langle \bar B_q| H^{\Delta
F=2}_{\rm eff}| B_q \rangle|$.
Explicitly, we have
\begin{eqnarray}
\left(\Delta m_{B_q}\right)_{X_P}&=&\frac{4g^2_Xm_{B_q}
f^2_{B_q}}{3(m^2_{B_q}-m^2_{X_P})}  \left|P^{SLL}_{1}-P^{LR}_{1}\right|\,, \nonumber \\
\left(\Delta m_{B_q}\right)_{X_A}&=&\frac{4g^2_Xm_{B_q}
f^2_{B_q}}{3(m^2_{B_q}-m^2_{X_A})}
\left|-(P^{VLL}_{1}-P^{LR}_{2})+\frac{(m_{b}+m_{q})^2}{m^2_{X_A}}
\left(P^{SLL}_{1}-P^{LR}_{1}\right)\right|\, , \label{eq:massdif}
\end{eqnarray}
where we have used the hadronic matrix elements, defined by \cite{BJU_NPB}
\begin{eqnarray}
\langle \bar B_q| (\bar b \ga^{\mu} P_{L(R)} q)(\bar b
\ga_{\mu}P_{L(R)} q)|
B_q\rangle&=&\frac{m_{B_q} f^2_{B_q}}{3}P^{VLL}_{1}\,, \nonumber \\
\langle \bar B_q| (\bar b \ga^{\mu} P_{L} q)(\bar b \ga_{\mu} P_{R}
q)|
B_q\rangle&=&\frac{m_{B_q} f^2_{B_q}}{3}P ^{LR}_{1}\,, \nonumber \\
\langle \bar B_q| (\bar b P_{L(R)} q)(\bar b P_{L(R)} q)|
B_q\rangle&=&\frac{m_{B_q} f^2_{B_q}}{3}P^{SLL}_{1}\,, \nonumber \\
\langle \bar B_q| (\bar b P_{L} q)(\bar b P_{R} q)|
B_q\rangle&=&\frac{m_{B_q} f^2_{B_q}}{3}P^{LR}_{2}\,, \label{eq:PPh}
\end{eqnarray}
with $f_{B_q}(m_{B_q})$ being the $B_q$ decay constant (mass).
It is clear that by using
$s(d)$ instead of $b(q)$,  Eq.~(\ref{eq:PPh}) can be
applied to the $K$ system. In
Table~\ref{table:PPbar} \cite{BJU_NPB},
we give the values of $P^{VLL}_{1}$,
$P^{LR}_{1,2}$ and $P^{SLL}_{1}$ for K and B systems. Note that
$P^{LR}_{1,2}$ and $P^{SLL}_{1}$ in the $B_q$ system are
much smaller than those in the $K$ system as
there
is no enhancement from chiral symmetry breaking in $B_{q}$.
\begin{table}[hptb]
\caption{The form factors for the $P-\bar P$ transitions with P
being the pseudoscalar mesons. }\label{table:PPbar}
\begin{ruledtabular}
\begin{tabular}{ccccc}
$P$ & $P^{VLL}_{1}$ & $P^{LR}_{1}$& $P^{SLL}_{1}$ & $P^{LR}_{2}$ \\ \hline 
 $K$ &  $0.48$ & $-36.1 $ & $-18.1$ &  $59.3$   \\ \hline
 $B_{q}$ & $0.84$ & $-1.62$ & $-1.47$ & $2.46$ \\
 \end{tabular}
\end{ruledtabular}
\end{table}

We now examine
$P\to \mu^{+} \mu^{-}$ with
$P=(K_{L},\, B_q)$.
Note that the decay of $K_{L}\to \mu^{+} \mu^{-}$ gives the strongest constraint on $\gq_{sd}$.
To estimate the decay BRs,
we use
\begin{eqnarray}
\langle 0| \bar q_{i} \ga_{\mu} \g5 q_{j} | P(p)\rangle &=& if_{P}
p_{\mu} \,,\ \ \
\langle 0| \bar q_{i}  \g5 q_{j} | P(p)\rangle =
-i\frac{f_{P}m^2_{P}}{m_{q_i}+m_{q_j}} \, .
\end{eqnarray}
By neglecting the mixing induced CP violation in the K system and using
$K_{L}\approx K_2$,
the X-mediated
decay  amplitudes for $P\to \mu^{+} \mu^{-}$   could be
summarized as
\begin{eqnarray}
M(P\to \mu^{+} \mu^{-})_{X_P}&=& -i a_{P}\frac{ \gq_{Pij}
\gl_{P\mu}}{m^2_{P} -m^2_{X_P}}\frac{f_{P}
m^{2}_{P}}{m_{q_i}+m_{q_j}} \bar
\mu \g5 \mu\, , \nonumber\\
M(P\to \mu^{+} \mu^{-})_{X_A}&=& i2a_{P}\frac{\gq_{Pij}
\gl_{P\mu}}{m^2_{X_A}}m_{\mu} f_{P} \bar \mu \g5 \mu\,,
\end{eqnarray}
where $q_i=(s,\, b)$, $q_{j}=(d,\, q)$ and $a_{P}=(\sqrt{2},1)$
for $P=(K_L, B_{q})$.
The corresponding decay
rates are given by
\begin{eqnarray}
\Gamma(P\to \mu^{+} \mu^{-})_{X_P}&=& \frac{m_{P}}{m_{X_P}}\left|
\frac{a_{P} \gq_{Pij} }{m^2_{P} -m^2_{X_P}}\frac{f_{P}
m^{2}_{P}}{m_{q_i}+m_{q_j}}\right|^{2}\frac{\sqrt{1-4m^2_{\mu}/m^2_{P}}
}{\sqrt{1-4m^2_{\mu}/m^2_{X_P}} }\Gamma(X_{P}\to \mu^{+} \mu^{-})\,,\nonumber \\
\Gamma(P\to \mu^{+} \mu^{-})_{X_A}&=&
\frac{3}{2}\frac{m_{P}}{m_{X_{A}}}\left| \frac{2a_{P}\gq_{Aij}
}{m^2_{X_A}}m_{\mu} f_{P}\right|^2\frac{\sqrt{1-4m^2_{\mu}/m^2_{P}}
}{(1-4m^2_{\mu}/m^2_{X_A})^{3/2} }\Gamma(X_{A}\to \mu^{+} \mu^{-})\,,
\label{eq:rate_2mu}
\end{eqnarray}
with%
\begin{eqnarray}
\Gamma(X_{P}\to \mu^{+} \mu^{-})&=&\frac{|\gl_{P\mu}|^2
m_{X_P}}{8\pi}
\sqrt{1-\frac{4m_{\mu}^2}{m^2_{X_P}}}\,,\nonumber \\
\Gamma(X_{A}\to \mu^{+} \mu^{-})&=&\frac{|\gl_{A\mu}|^2
m_{X_A}}{12\pi} \left(1-\frac{4m_{\mu}^2}{m^2_{X_A}}\right)^{3/2}.
\label{eq:rate}
\end{eqnarray}

After introducing $|\Delta F|=2$ ($F=S,B$) and purely rare
dileptonic decays of $K_{L}$ and $B_{q}$, we investigate other
X-mediated rare semileptonic processes, such as $B_{q}\to M X\to
M\mu^{+} \mu^{-}$ with $M$ being a light meson.
For the s-wave states in the processes,
$M$ has to be a vector-meson $(V)$ due to
parity. On the other hand,
for the production
of p-wave states in the  decays, $M$
can be either scalar $(S)$ or axial-vector $(A)$ mesons, such as $f_0(980)$,
$a_{0}(980)$, $\kappa$, $f_{0}(1500)$, $K^{*}_{0}(1430)$ and $K_{1}$.
To study
the decay rates,
 we illustrate the formulas in
the case with $M=V$ and $X=X^{\mu}_{A}$.
We write
the corresponding decay amplitude to be
\begin{eqnarray}
M=\langle V \ell^{+} \ell^{-} | {\cal H}_{\rm eff}| B_{q}\rangle =
h_{\mu} \frac{-g^{\mu\nu} + q^{\mu}
q^{\nu}/m^2_{X_{A}}}{q^2-m^2_{X_{A}}+i\Gamma_{X_{A}} m_{X_{A}}} \bar
\mu \ga_{\nu} \g5 \mu\,,
\end{eqnarray}
where $h_{\mu}=\langle V|\bar b \ga_{\mu}\g5 \qp|B_{q}\rangle$
denotes the hadronic transition matrix element and $\Gamma_{X_A}$ is
the total decay width of $X^{\mu}_{A}$. In terms of the narrow width
approximation, given by
\begin{eqnarray}
\left|\frac{1}{q^2-m^2_{X_{A}}+i\Gamma_{X_{A}} m_{X_{A}}}\right|^2
\approx \frac{\pi}{\Gamma_{X_{A}}
m_{X_{A}}}\delta(q^2-m^2_{X_{A}})\, ,
\end{eqnarray}
the squared decay amplitude can be written as
 \begin{eqnarray}
|M|^{2}&=& \left|\sum_{\la=0,\pm} h_{\mu}
\ve^{\mu*}_{X_{A}}(\la) \ve^{\nu}_{X_{A}}(\la) \bar \mu \ga_{\nu}\g5
\mu \right|^2 \frac{\pi}{\Gamma_{X_{A}}
m_{X_{A}}}\delta(q^2-m^2_{X_{A}}) \, ,
\nonumber \\
&=& \left[\sum_{\la=0,\pm} \left|h_{\mu} \ve^{\mu*}_{X_{A}}(\la)
\right|^2 \right] \left[ \sum_{\la=0,\pm} \left|
\ve^{\nu}_{X_{A}}(\la) \bar \mu \ga_{\nu}
\g5\mu\right|^2\right]\frac{\pi}{\Gamma_{X_{A}}
m_{X_{A}}}\delta(q^2-m^2_{X_{A}})\, , \label{eq:mm}
  \end{eqnarray}
%
and the
differential decay rates for the semileptonic $B_q$ decays
are given by
\begin{eqnarray}
d\Gamma(B_{q}\to V \mu^{+}
\mu^{-})
&=&
 \frac{1}{2m_{B_q}}\left[\sum_{\la=0,\pm} \left|h_{\mu}
\ve^{\mu*}_{X_{A}}(\la) \right|^2 \right]d_{2}(B_{d}\to V X_{A})
\nonumber\\
&&\times
 \frac{1}{2m_{X_{A}} \Gamma_{X_{A}}} \left[ \sum_{\la=0,\pm}
\left| \ve^{\nu}_{X_{A}}(\la) \bar \mu \ga_{\nu}
\g5\mu\right|^2\right]d_{2}(X_{A}\to \mu^{+} \mu^{-})\, \nonumber \\
&=& 3 d\Gamma(B_{q}\to V X_{A})\, dBR(X_{A}\to \mu^{+} \mu^{-})\,,
\label{dG}
\end{eqnarray}
where the factor $3$ is due to the spin-degree of freedom from $X_{A}$. The formula in Eq. (\ref{dG})
could be applied to any decaying chain of $P\to M X\to
M\mu^{+} \mu^{-}$.
As a result, we find that
\begin{eqnarray}
\Gamma(B_{q}\to MX \to M\mu^{+}
\mu^{-})=f_{X} \Gamma(B_{q}\to M X)\times BR(X\to \mu^{+} \mu^{-})\,,
\end{eqnarray}
where $f_{X}=(1,3)$ for
$X=(X_{P},X_{A})$, representing the spin degree of freedom.

In our following analysis, we will
focus on the decays of $B_{q}\to MX$. It is clear that
$X$ is emitted and the meson in the final state owns the same
spectator quark as $B_q$. The hadronic effects for the decays are
only related to $B_{q}\to M$ transition form factors.
The
relevant form fcators for various mesons are parametrized by \cite{LF}
\begin{eqnarray} \langle
V(p,\ve_{V})| \bar b \ga^{\mu} \g5 q^{\prime}|
B_{q}(p_{B})\rangle&=&i\left[(m_{B_q}-m_{V})\left(
\ve^{\mu*}_{V}(\la) -\frac{\ve^{*}_{V}(\la)\cdot
P}{q^2}q^{\mu}\right) A^{B_q}_{1} \right.
\nonumber\\
&& -\frac{\ve^{*}_{V}\cdot P}{m_{B_q}+m_{V}}
\left(P^{\mu}-\frac{P\cdot q}{q^2}q^{\mu} \right)A^{B_q
V}_{2}(q^2)\nonumber \\
&&\left. +2m_{V}\frac{\ve^{*}\cdot P}{q^2}q^{\mu} A^{B_q V}_{0}(q^2)
\right]\,, \nonumber \\
\langle A(p,\ve_{A})| \bar b \ga_{\mu} \g5 q^{\prime}|
B_{q}(p_B)\rangle&=& -\frac{A^{B_q
A}(q^2)}{m_{B_q}-m_{A}}\epsilon_{\mu\nu\rho
\sigma }\ve^{\nu*}_{A}(\la) P^{\rho} q^{\sigma}\,, \nonumber \\
\langle S(p)|\bar b \ga^{\mu} \g5 q^{\prime}| B_{q}(p_{B})\rangle
&=& -i\left[ \left(P_{\mu}-\frac{P\cdot q}{q^2}q_{\mu} \right)F^{B_q
S}_{1}(q^2)+\frac{P\cdot q}{q^2}q_{\mu} F^{B_q S}_{0}(q^2)
\right]\,, \label{eq:ff}
\end{eqnarray}
where $\ve_{V(A)}(\la)$ denotes the polarization of $V(A)$
with the helicity state $\lambda$,
$P_{\mu}=(p_{B}+p)_{\mu}$ and $q_{\mu}=(p_{B}-p)_{\mu}$.
 Consequently,
 the decay amplitudes for the decays $B_{q}\to (V,\, S)
X_{P}$ can be written as
 \begin{eqnarray}
 A(B_{q}\to V X_{P})&=& -\gq_{Pbq^{\prime}}\frac{2m_{V}}{m_{b}+m_{q^{\prime}}
 }\ve^{*}_{V}\cdot q A^{B_q V}_{0}(m^2_{X_P})\,, \nonumber\\
A(B_{q}\to S
X_{P})&=&\gq_{Pbq^{\prime}}\frac{m^{2}_{B_q}-m^{2}_{S}}{m_{b}+m_{q^{\prime}}
 }F^{B_q
S}_{0}(m^2_{X_P})\,.
\label{ABVS}
 \end{eqnarray}
Since $m_{X}$
is as light as $0.214$ GeV, it is
a good approximation to adopt $F(0)\approx F(m_X^2)$ for the
various
form factors.
 From Eq. (\ref{ABVS}), we obtain
 \begin{eqnarray}
\Gamma(B_{q}\to V X_{P})&\approx & \frac{m^3_{B_q} \left(A^{B_q
V}_{0}(0)\right)^2}{16\pi (m_{b}+m_{\qp})^2} \left(
1-\frac{m^2_{V}}{m^2_{B_q}}\right)^3 \left|\gq_{Pbq^{\prime}}
\right|^2\, ,\nonumber \\
\Gamma(B_{q}\to S X_{P})&\approx & \frac{m^3_{B_q} \left(F^{B_q
S}_{0}(0)\right)^2}{16\pi (m_{b}+m_{\qp})^2} \left(
1-\frac{m^2_{S}}{m^2_{B_q}}\right)^3 \left|\gq_{Pbq^{\prime}}
\right|^2\,. \label{eq:brxp}
 \end{eqnarray}
Similarly, the decay amplitudes for $B_{q}\to (V,\,A,\, S) X_{A}$
 are given by
 \begin{eqnarray}
A(B_{q}\to V X_{A})&=& i\gq_{Abq^{\prime}}\left[ (m_{B_q}+m_{V})
A^{BV}_{1}(0)
\ve^{*}_{V}(\la)\cdot \ve^{*}_{X_A}(\la)\right.\nonumber\\
&&\left. -\frac{2A^{BV}_{2}(0)}{m_{B_q}+m_{V}} \ve^{*}_{V}(\la)
\cdot p_{B}\;
\ve^{*}_{X_A}(\la)\cdot p_{B}\right]\,, \nonumber\\
A(B_{q}\to A X_{A})&=& -\frac{2\gq_{Ab\qp}
A^{B_qA}(0)}{m_{B_q}-m_{A}} \epsilon_{\mu \nu \rho \sigma}
\ve^{\mu*}_{X_A}(\la) \ve^{\nu*}_{A}(\la) p^{\rho}_{B} q^{\sigma}\,, \nonumber\\
A(B_{q}\to S X_{A})&=&-2i\gq_{Abq^{\prime}} \ve^{*}_{X_A}(\la) \cdot
p_{B}F^{B_q S}_{1}(0)\,. \label{eq:ampvv}
 \end{eqnarray}
To
get the unified formulas for the decay rates with two spin-1 mesons
in the final states,
we can write the general decay amplitude
in terms of the helicity basis as
\begin{eqnarray}
A_{\la}=\ve^{*}_{1\mu}(\la) \ve^{*}_{2\nu}(\la)\left[ag^{\mu \nu} +
\frac{b}{m_{1} m_{2}} p^{\mu}_{B} p^{\nu}_{B} +i\frac{c}{m_1 m_2}
\epsilon^{\mu \nu \alpha \beta}p_{1\alpha} p_{2\beta}
\right]\label{eq:hamp}
\end{eqnarray}
where $m_{1,2}$ stand for the meson masses. Then, the longitudinal
and transverse components are given by $H_{0}=-ax-b(x^2-1)$ and
$H_{\pm}=a\pm c\sqrt{x^2-1}$, respectively, with
$x=(m^2_B-m^2_{1}-m^2_{2})/(2m_1 m_2)$. Since $B_{q}$ is a spinless
particle, both spin-1 mesons in the final states should have the
same helicity.
Hence, the decay rates are given by
\cite{KP,Chen_1}
 \begin{eqnarray}
  \Gamma(B_{q}\to M_1 M_2)&=& \frac{|\vec{p}|}{8\pi m^2_{B_q}}
  \left(|H_{0}|^2 +|H_{+}|^2 +|H_{-}|^2 \right)\label{eq:brvv}
 \end{eqnarray}
with  $|\vec{p}|=m_1 m_{2}\sqrt{x^2-1}/m^2_{B_q}$.
By comparing
Eq.~(\ref{eq:ampvv}) with Eq.~(\ref{eq:hamp}), we find that
\begin{eqnarray}
a=i\gq_{Ab\qp}(m_{B_q}+m_{V}) A^{B_qV}_{1}(0)\,, \ \ \
b=-i\gq_{Ab\qp}m_{V}m_{X_{A}}\frac{2A^{B_q
V}_{2}(0)}{m_{B_q}+m_{V}}\, , \ \ \ c=0\,,
\end{eqnarray}
for
$B_q\to V X_{A}$ and
 \begin{eqnarray}
 a=0\,\ \ \ b=0\,, \ \ \ c=i\gq_{Ab\qp}m_{A}m_{X_{A}} \frac{2A^{B_q
 A}(0)}{m_{B_q}-m_{A}}\,,
 \end{eqnarray}
for $B_{q}\to A X_{A}$.
The decay rates for $B_{q}\to S X_{A}$ are given by
\begin{eqnarray}
\Gamma(B_q\to S X_{A})&\approx &\frac{m^3_{B_q}}{16\pi}
\left(\frac{F^{B_q S}_{1}(0)}{m_{X_A}}\right)^2 \left(
1-\frac{m^2_{S}}{m^2_{B_q}}\right)^3 \left|\gq_{Ab\qp} \right|^2\, .
\end{eqnarray}

Finally,  we study the rare decays of $B_{q}\to \ga
X_A$. Although $\ga$ is a vector boson, unlike an ordinary vector
meson, it is massless and only has transverse degrees of freedom.
With the transition
form factor for $B_{q}\to \ga$, defined by \cite{GLZ_PRD}
\begin{eqnarray}
\langle \ga(k)| \bar b \ga_{\mu} \g5
q^{\prime}|B_{q}(p_{B})\rangle&=& ie \frac{F_{A}(q^2)}{m_{B}} \left[
\ve^{\mu*}_{\ga}(\la) p_{B}\cdot k-\ve^{*}_{\ga}(\la)\cdot p_{B}
k_{\mu}\right]\,,
\end{eqnarray}
the decay amplitudes are given by
\begin{eqnarray}
A(B_{q}\to \ga X_{A})&=& ie\frac{F_{A}(q^2)}{m_{B_q}} p_{B}\cdot k
\ve^{*}_{\ga}(\la )\cdot \ve^{*}_{X_A}(\la) \,,
\end{eqnarray}
leading to
\begin{eqnarray}
\Gamma(B_{q}\to \ga X_{A})=\frac{\alpha_{\rm em}}{8}
m_{B_q}|F_{A}(0)|^2 \left|\gq_{Abq^{\prime}}\right|^2\label{eq:gaxa}
\end{eqnarray}
with $\alpha_{\rm em}=e^2/4\pi$.

After  the $B_{q}$ processes,  we now discuss
the pure leptonic decays.
As the new particle of $X_{P}$ or $X^{\mu}_{A}$
couples to the muon, it is natural to
speculate that it
 will also
couple to other leptons such as $e$ and $\tau$.
Moreover, the new particle
could also give rise to the LFV, like the FCNCs in the quark sector,
such as $\mu\to e \ga$, $\mu\to 3e$ and $\tau\to \ell
\mu^{+} \mu^{-}$. Recently, the LFV in $\tau$ decays has been
improved up to $O(10^{-7})$ \cite{belle-tau,babar-tau}.
It should be interesting to explore the LFV due to $X_{P,A}$.
 To illustrate the effects of the LFV,
 we will concentrate on the processes related to $\tau$.
We write the relevant effective interactions
as
\begin{eqnarray}
-{\cal L}=-i\gl_{P\ell \tau}\bar \ell \g5 \tau X_{P} + \gl_{A\ell \tau}
\bar \ell \ga_{\mu} \g5 \tau X^{\mu}_{A} +H.c.\,. \label{eq:tau}
\end{eqnarray}
The rates for $\tau\to \ell X\to
\ell \mu^{+} \mu^{-}$ are described by $\Gamma(\tau\to \ell
X\to \ell \mu^{+} \mu^{-})= f_{X} \Gamma(\tau\to \ell X)BR(X\to
\mu^{+} \mu^{-})$.
Similar to the $B_{q}$ decays, we only discuss
$\tau\to \ell X$. In terms of the interactions
in Eq.~(\ref{eq:tau}),
the decay rates with $X_{P}$ and $X_{A}$ are given by
\begin{eqnarray}
\Gamma(\tau\to \ell X_{P})&\approx& \frac{m_{\tau}}{16\pi} \left|
\gl_{P\ell\tau}\right|^2\, ,\ \ \  \Gamma(\tau\to \ell X_{A})\approx
\frac{m^3_{\tau}}{32\pi m^2_{X_A}} \left|
\gl_{P\ell\tau}\right|^2\,,\label{eq:rate_tau}
\end{eqnarray}
respectively,
where we have neglected the masses of $\ell$ and $X_{P,A}$ due to
$m_{\tau}\gg m_{X_{P,A}}> m_{\ell}$.

\begin{table}[hptb]
\caption{The input values
of
parameters in units of $GeV$.
}\label{table:values}
\begin{ruledtabular}
\begin{tabular}{ccccccc}
$f_{K}$ & $f_{B_d}$ & $f_{B_s}$& $m_{X}$ & $m_{K^0}$ & $m_{B_d}$ & $m_{B_s}$ \\ \hline 
$0.16$ &  $0.20$ & $0.22 $ & $0.214$ &  $0.497$ & $5.28$ & $5.37$
\\\hline\hline
$m_{s}$ & $m_{d}$ & $m_{b}$ & $\tau_{K_{L}}$ & $\tau_{B_d}$ &
$\tau_{B_s}$ &
\\ \hline
$0.15$ & $0.01$ & $4.4$ & $7.87\times 10^{16}$ & $2.33\times 10^{12}$ & $2.22\times 10^{12}$ &
\\
\end{tabular}
\end{ruledtabular}
\end{table}
In order to do the numerical estimations, the input values for the
various parameters are presented in Table~\ref{table:values}. To
see the effects of the new particle on low energy physics, we
first consider its contributions to $\Delta F=2$ processes.
>From the current experimental data,
the mass
differences in the $K$ and $B_q$ systems are given by
$ \Delta m_{K}= (3.483 \pm
0.006)\times 10^{-15}$, $\Delta m_{B_d}= (3.337\pm 0.033)\times
10^{-13}$ and $\Delta m_{B_s}=(11.45^{+0.20}_{-0.13})\times
10^{-12}$ GeV.
By utilizing these values and those inputs in
Tables~ \ref{table:PPbar} and \ref{table:values}, from
Eq.~(\ref{eq:massdif}) the direct constraints on the couplings
are found to be
\begin{eqnarray}
\left| \gq_{Psd}\right|^2 &<& 2.3 \times 10^{-15}\,,\ \ \ \left|
\gq_{Pbd}\right|^2< 2.2\times 10^{-10}\,,\ \ \
 \left|
\gq_{Pbs}\right|^2 < 6.4 \times 10^{-9}\,, \nonumber \\
\left| \gq_{Asd}\right|^2&<& 0.67 \times 10^{-15}\,, \ \ \ \left|
\gq_{Abd}\right|^2< 5.1\times 10^{-13}\,\ \ \ \left|
\gq_{Abs}\right|^2< 1.4\times 10^{-11}. \label{eq:limit_dm}
\end{eqnarray}
Due to the strong cancelation between $P^{SLL}_{1}$ and $P^{LR}_{1}$
in $B_q$, the constraints from $\Delta m_{B_q}$ are two orders of
magnitude less than the naive expectation.
Next, we discuss the decays of
$P\to
\mu^{+} \mu^{-}$. It is well known that the long-distance effect
dominates the process of $K_{L}\to \mu^{+} \mu^{-}$, while the
short-distance contribution usually is taken to be $BR(K_{L}\to
\mu^{+} \mu^{-})_{SD}<3.6\times 10^{-10}$ \cite{HTV_PLB}. As to the
dileptonic decays in $B_q$ decays, we also know the upper bounds of
$BR(B_d\to \mu^{+} \mu^{-})<2.3 \times 10^{-8}$ and $BR(B_{s}\to
\mu^{+} \mu^{-})< 8\times 10^{-8}$ \cite{HFAG}. With these
constraints and  Eq.~(\ref{eq:rate_2mu}), we have
\begin{eqnarray}
\left| \gq_{Psd}\right|^2 \Gamma(X_{P}\to \mu^{+} \mu^{-})&<&
6.5\times 10^{-29}\,,\ \ \ \left| \gq_{Pbd}\right|^2 \Gamma(X_{P}\to
\mu^{+} \mu^{-}) < 3.8\times 10^{-20}\,,\nonumber \\
 \left|
\gq_{Pbs}\right|^2 \Gamma(X_{P}\to \mu^{+} \mu^{-})&<& 1.2 \times
10^{-19}\,, \ \ \ \left| \gq_{Asd}\right|^2 \Gamma(X_{A}\to \mu^{+}
\mu^{-})< 1.0 \times 10^{-29}\,,
\nonumber \\
 \left| \gq_{Abd}\right|^2 \Gamma(X_{A}\to \mu^{+} \mu^{-})&<& 2.3 \times 10^{-24}\,\ \ \
\left| \gq_{Abs}\right|^2 \Gamma(X_{A}\to \mu^{+} \mu^{-})<
7.0\times 10^{-24}. \label{eq:limit_b2mu}
\end{eqnarray}
It has been shown that the current strict bounds on $|\gl_{P\mu}|^2$
and $|\gl_{A\mu}|^2$ are from muon $g-2$, given by
$|\gl_{P\mu}|^2<2.6 \times 10^{-7}$ and $|\gl_{A\mu}|^2<6.7 \times
10^{-8}$ \cite{HTV_PLB}, respectively.
>From Eq.~(\ref{eq:rate}), one gets the
upper bounds on the rates as $\Gamma(X_{P}\to \mu^{+} \mu^{-})< 4.3
\times 10^{-10}$ GeV and $\Gamma(X_{A}\to \mu^{+} \mu^{-}) < 2.7
\times 10^{-12}$ GeV. To illustrate the constraints on the couplings,
we take  $\Gamma(X_{P,A}\to \mu^{+} \mu^{-})\sim
 (10^{-10}, 10^{-12})$ GeV and we obtain
%
\begin{eqnarray}
\left| \gq_{Psd}\right|^2 &<& 6.5 \times 10^{-19}\,,\ \ \ \left|
\gq_{Pbd}\right|^2< 3.8\times 10^{-10}\,,\ \ \
 \left|
\gq_{Pbs}\right|^2 < 1.2 \times 10^{-9}\,, \nonumber \\
\left| \gq_{Asd}\right|^2&<& 1.0 \times 10^{-17}\,, \ \ \ \left|
\gq_{Abd}\right|^2< 2.3 \times 10^{-12}\,\ \ \ \left|
\gq_{Abs}\right|^2< 7.0\times 10^{-12}, \label{eq:limit_dm}
\end{eqnarray}
respectively.
It is clear that the constraints from $\Delta m_{K}$
are weaker than those from $K_L\to \mu^{+} \mu^{-}$, which
are consistent with
 the HyperCP data in the decay of $\Sigma^{+}\to
p^{+}\mu^{+} \mu^{-}$,
given by
\cite{HTV_PLB}
\begin{eqnarray*}
\left| \gq_{Psd}\right|^2 BR(X_{P}\to \mu^{+}
\mu^{-})&=& (8.4^{+6.5}_{-5.1}\pm 4.1)\times 10^{-20}\,,\nonumber \\
\left| \gq_{Asd}\right|^2 BR(X_{A}\to \mu^{+} \mu^{-})&=&
(4.4^{+3.4}_{-2.7}\pm 2.1)\times 10^{-20}\,.
\end{eqnarray*}
On the other hand, the
bounds from $\Delta m_{B_q}$ and $BR(B_{q}\to \mu^{+} \mu^{-})$ are
similar.

 To estimate
the BRs of semileptonic $B_q$ decays, we use
the $B_{q}\to M$ transition form factors in
Eq.~(\ref{eq:ff}),
calculated by the light-front quark model (LFQM) and
summarized in
Table~\ref{table:vff} \cite{LF}.
\begin{table}[hptb]
\caption{Values of form factors at $q^2=0$ defined in
Eq.~(\ref{eq:ff}) and calculated by the LFQM \cite{LF}. }\label{table:vff}
\begin{ruledtabular}
\begin{tabular}{ccccccc}
$A^{B_q K^*}_{0} $ & $A^{B_q K^*}_{1}$ & $A^{B_q K^*}_{2}$ & $A^{B_q
K_{^3P_1}}$ &
$A^{B_q K_{^1P_1}}$& $F^{B_q K^{*}_{0}}_{1}$ & $F^{B_q K^{*}_{0}}_0$ \\ \hline 
$0.31$ &  $0.26$ & $0.24$ & $0.26$ &  $0.11$ & $0.26$ & $0.26$ \\
 \end{tabular}
\end{ruledtabular}
\end{table}
In the table, the states of $^{1}P_{1}$ and $^{3}P_{1}$ will be used to
consist of the physical states $K_1(1270)$ and $K_{1}(1400)$ and
their relations are parametrized by \cite{LF,Suzuki},
 \begin{eqnarray}
 K_1(1270)=K_{^1\!P_1}\cos\theta + K_{^3P_1} \sin\theta\,,
 \nonumber\\
 K_1(1400)=-K_{^1\!P_1}\sin\theta+K_{^3P_1} \cos\theta\,,
 \label{eq:mixing}
 \end{eqnarray}
with $\theta=58^{\circ}$ being the mixing angle \cite{Chen_1}.
The decay of $B_{d}\to
K^{*0} \mu^{+} \mu^{-}$ has been measured at the B-factories with
 the world average on the decay BR
 being
$(1.22^{+0.38}_{-0.32})\times
10^{-6}$ \cite{PDG_06}.
>From Eqs.~(\ref{eq:brxp}), (\ref{eq:ampvv})
and (\ref{eq:brvv}) and the values in Tables
\ref{table:values} and \ref{table:vff},
we obtain
\begin{eqnarray*}
BR(B_{d}\to K^{*0} X_{P}\to K^{*0} \mu^{+} \mu^{-})&=& 3.1\times
10^{10}\left| \gq_{Pbs}\right|^2 BR(X_{P}\to \mu^{+} \mu^{-})\,,
\nonumber \\
BR(B_{d}\to K^{*0} X_{A}\to K^{*0} \mu^{+} \mu^{-})&=& 3.9\times
10^{13}\left| \gq_{Abs}\right|^2 BR(X_{A}\to \mu^{+} \mu^{-})\,.
\end{eqnarray*}
If we regard $BR(B_{d}\to K^{*0} \mu^{+} \mu^{-})=1.22\times
10^{-6}$ as the upper bound, we have
\begin{eqnarray}
 \left| \gq_{Pbs}\right|^2 BR(X_{P}\to \mu^{+} \mu^{-})&\leq & 3.9
\times 10^{-17}\,,
\nonumber \\
 \left| \gq_{Abs}\right|^2 BR(X_{A}\to \mu^{+} \mu^{-})&\leq & 3.1\times
 10^{-20}\,. \label{eq:bound_q}
\end{eqnarray}
For
$BR(X_{P,A}\to \mu^{+} \mu^{-})
\sim 1$,
 we find  that the decay $B_{d}\to K^{*0} \mu^{+}
\mu^{-}$ gives the strongest limits on the couplings of $\gq_{Pbs}$
and $\gq_{Abs}$.

 From Eq. (\ref{eq:bound_q}),
 we can
study the contributions of $X_{P,A}$ to other $B_{q}$ decays. The
first direct
application is $B_s\to \phi X_{P,A}\to \phi \mu^{+} \mu^{-}$. In
terms of the formulas shown in Eqs.~(\ref{eq:brxp}),
(\ref{eq:ampvv}) and $(\ref{eq:brvv})$, we get
\begin{eqnarray}
BR(B_s\to \phi X_{P}\to \phi \mu^{+} \mu^{-}) &\leq & 2.74 \times
10^{-6}\,, \nonumber\\
BR(B_s\to \phi X_{A}\to \phi \mu^{+} \mu^{-}) &\leq & 2.81 \times
10^{-6}\,,
\end{eqnarray}
where we have used $A^{B_s\phi}_{0}(0)=0.474$,
$A^{B_s\phi}_{1}(0)=0.311$ and $A^{B_s\phi}_{2}(0)=0.234$ for
$B_s\to \phi$ transition form factors calculated by the light cone sum
rules (LCSRs) \cite{BW}. Interestingly, the bounds are just
under the D0
upper limit of
$BR(B_s\to \phi \mu^{+} \mu^{-})<
3.2 \times 10^{-6}$ \cite{D0}. If the events observed by
the HyperCp collaboration \cite{Park}
are indeed from
the new particle,
 the decay
of $B_s\to \phi \mu^{+} \mu^{-}$ should be observed soon as the
standard model prediction is around $1.6\times 10^{-6}$ \cite{GL}.

Next, we discuss the productions of p-wave mesons in $B_q$ decays. As
mentioned before, the p-wave mesons could be $f_{0}(980)$,
$a_{0}(980)$, $\kappa$, $K^{*}_{0}(1430)$ and $K_{1}$. However,
since the quark contents for the light p-wave mesons ($<$1 GeV) are
not certain, we will only focus on $K^{*}_{0}(1430)$ and $K_{1}$. By
using the values given in Tables~\ref{table:values} and
\ref{table:vff}, we directly display the predicted upper BRs in
Table~\ref{table:brs}. From the table, we see clearly that only the
production of $K^{*}_{0}(1430)$ is interesting, which is accessible
to the current B-factories. As only the transverse degrees of
freedom are involved in the decays of  $B_{q}\to A X_{A}$, where the
effects are proportional to the masses of $A$ and $X_A$, one  can
easily understand  why the BRs for $B_q\to K_{1} X_{A}\to K_{1}
\mu^{+} \mu^{-}$ are so small and negligible.
\begin{table}[hptb]
\caption{ BRs (in units of $10^{-6}$) for $B_d\to (K^{*}_{0}(1430),\,
K_{1}(1270),\, K_{1}(1400))X_{P,A}(X_{P,A}\to \mu^{+}
\mu^{-})$.}\label{table:brs}
\begin{ruledtabular}
\begin{tabular}{ccc}
Mode & $B_{d}\to K^{*}_{0}(1430)X_P(X_P\to\mu^{+} \mu^{-})$ &
$B_{d}\to
K^{*}_{0}(1430)X_A(X_A\to\mu^{+} \mu^{-})$   \\ \hline 
BR &  $0.74$ & $0.75$  \\ \hline\hline 
Mode & $B_{d}\to K_{1}(1270)X_A(X_A\to\mu^{+} \mu^{-})$ & $B_{d}\to
K_{1}(1400)X_A(X_A\to \mu^{+} \mu^{-})$ \\ \hline 
BR &  $6\times 10^{-3}$ & $1.54\times 10^{-4}$   \\
 \end{tabular}
\end{ruledtabular}
\end{table}
In addition, according to Eq.~(\ref{eq:gaxa}), we also get that
$BR(B_s\to \ga X_{A}\to \ga\mu^{+} \mu^{-})$ is less than
$5.4\times 10^{-11}$, which is much smaller than the contributions
of in the SM \cite{GLW} and negligible.

As there is no any useful information on
$\gq_{P(A)bd}$, one can only investigate those decays
associated with $\gq_{P(A)bs}$.
In order to apply
$\gq_{P(A)bs}$ to decays related to $\gq_{P(A)bd}$, we need
some theoretical ansaz to connect them.
One of the interesting ansaz is to relate the couplings with quark masses,
such as
\begin{eqnarray}
\gq_{Pij}= \lambda_{P} \left(\frac{m_i m_j}{\rm v^2}\right)^{1\over 2}
\,, \ \ \ \gq_{Aij}= \lambda_{A} \left(\frac{m_i m_j}{\rm
v^2}\right)^{1\over 2} \,, \label{eq:ansaz_q}
\end{eqnarray}
where $\lambda_{P(A)}={\rm v}/{\rm v_{F}}$ denotes the ratio of
electroweak scale (v) to the new scale ($\rm v_{F}$) associated with
 the new particle.
  By using the ansaz, we find that
$\gq_{P(A)bd}/\gq_{P(A)bs}\sim (m_d/m_s)^{1/2}\sim 0.26$
and the upper bounds
on BRs for some decays associated with $\gq_{P(A)bd}$ are given by
\begin{eqnarray}
BR(B_u\to \rho^{+} X_{P}\to \rho^{+}\mu^{+} \mu^{-})&=&6.3\times
10^{-8}\,, \nonumber
\\
BR(B_u\to \rho^{+} X_{A}\to \rho^{+}\mu^{+} \mu^{-})&=&6.4\times
10^{-8}\,, \nonumber
\\
BR(B_s\to K^{*0} X_{P}\to K^{*0} \mu^{+} \mu^{-})&=&5.1 \times 10^{-8}\,, \nonumber \\
BR(B_s\to K^{*0}_{0}(1430) X_{P}\to K^{*0}_{0}(1430)\mu^{+} \mu^{-})&=& 4.9\times 10^{-8}\,, \nonumber \\
BR(B_s\to K^{*0}_{0}(1430) X_{A}\to K^{*0}_{0}(1430) \mu^{+}
\mu^{-})&=& 5.0\times 10^{-8}\,,
\end{eqnarray}
where the  form factors are taken to be
$A^{B_d\rho}(0)=0.28$, $A^{B_d\rho}_{1}(0)=0.22$,
$A^{B_d\rho}_{2}=0.20$ \cite{LF} and $A^{B_s K^*}=0.247$ \cite{BW},
respectively.

Finally, we study $\tau\to \ell \mu^{+} \mu^{-}$. Here, we use the same
ansaz as that for the quark sector in Eq. (\ref{eq:ansaz_q}),
$i.e.$,
\begin{eqnarray}
\gl_{P\ell\tau}= \lambda_{P}
\left(\frac{m_{\ell} m_{\tau}}{\rm
v^2}\right)^{1\over 2}\,, \ \ \ \gl_{A\ell\tau}= \lambda_{A}\left(
 \frac{m_{\ell}
m_{\tau}}{\rm v^2}\right)^{1\over 2}\,.
\end{eqnarray}
By Eqs.~(\ref{eq:rate_tau}), (\ref{eq:bound_q}) and
(\ref{eq:ansaz_q}) and with $\tau_{\tau}=4.4 \times 10^{11}$ GeV,
the upper bounds on the flavor violating $\tau$ decays are estimated
to be
\begin{eqnarray}
BR(\tau\to e X_{P}\to e \mu^{+} \mu^{-} )&=&8.5\times 10^{-10}\,, \nonumber \\
BR(\tau\to \mu X_{P}\to \mu \mu^{+} \mu^{-} )&=& 1.7 \times 10^{-7}\,,  \nonumber \\
BR(\tau\to e X_{A}\to e \mu^{+} \mu^{-} )&=&7.0 \times 10^{-11}\,, \nonumber \\
BR(\tau\to \mu X_{A}\to \mu \mu^{+} \mu^{-} )&=& 1.4\times
10^{-8}\,.
\end{eqnarray}
Note that the current upper bounds for
$\tau\to \ell \mu^{+} \mu^{-}$ are  $2.0 \times 10^{-7}$
\cite{PDG_06}.

In sum, we have studied the implications of the HyperCP Data on
 flavor changing  $B$ and $\tau$ decays.
 We have given constraints on the effective couplings due to
 the new pseudoscalar and axial-vector bosons of $X_{P,A}$
 from experimental data in the $K$ and $B$ systems, respectively.
 We have pointed out that the strongest limits on $g^{Q}_{P(A)bs}$
 are from the decay of $B_d\to K^{*0}\mu^{+}\mu^{-}$.
 We have shown that the decay BR of $B_{s}\to\phi\mu^{+}\mu^{-}$
 can be as large as $2.7 (2.8)\times 10^{-6}$ through $X_{P(A)}$,
 which are larger than the prediction of $1.6\times 10^{-6}$
 in the SM and close to the experimental upper limit of
 $3.2\times 10^{-6}$. Furthermore, we have found that
 $BR(B_{d}\to K_{0}^{*}(1430){P(A)}\to
 K_{0}^{*}(1430)\mu^{+}\mu^{-})$ is about
 $7.4\ (7.5)\times 10^{-7}$, whereas other related modes are negligible.
 In addition, we have proposed an ansaz to relate the couplings with
 the fermion masses. Based on this ansaz, we have demonstrated
 that the decay BRs of $B_{u}\to \rho^{+}\mu^{+}\mu^{-}$,
  $B_{s}\to K^{*0}\mu^{+}\mu^{-}$ and $B_{s}\to K^{*0}_{0}(1430)\mu^{+}\mu^{-}$ can all be at the level of $10^{-8}$. In particular,
  we have shown that $BR(\tau\to\mu X_{P(A)}\to\mu\mu^{+}\mu^{-})=1.7\ (0.14)\times 10^{-7}$. \\

\noindent
{\bf Acknowledgments}

This work is supported in part by the National Science Council of
R.O.C. under Grant \#s:NSC-95-2112-M-006-013-MY2
 and NSC-95-2112-M-007-059-MY3.


\begin{thebibliography}{99}
\bibitem{Park} H. Park {\it et al.}  [HyperCP Collaboration], Phys. Rev. Lett. {\bf 94}, 021801 (2005).

\bibitem{Bergstrom} L.~Bergstrom, R.~Safadi and P.~Singer, Z. Phys. {\bf C37}, 281 (1988).
\bibitem{He}
X.~G.~He, J.~Tandean and G.~Valencia,
  Phys.\ Rev.\ D {\bf 72}, 074003 (2005).

\bibitem{HTV_PLB}X.~G. He, J. Tandean and G. Valencia, Phys. Lett.
B{\bf 631}, 100 (2005).

\bibitem{DEJ}
N.~G.~Deshpande, G.~Eilam and J.~Jiang,
  Phys.\ Lett.\ B {\bf 632}, 212 (2006).
\bibitem{GH} C.~Q.~Geng and Y.~K.~Hsiao,
  Phys.\ Lett.\ B {\bf 632}, 215 (2006).

\bibitem{GR}
D.~S.~Gorbunov and V.~A.~Rubakov,
  Phys.\ Rev.\ D {\bf 64}, 054008 (2001);
  Phys.\ Rev.\ D {\bf 73}, 035002 (2006);
 S.V. Demidov and D.S. Gorbunov, arXiv:hep-ph/0610066.

 \bibitem{HTV-LH}X.~G.~He, J.~Tandean and G.~Valencia,
 arXiv:hep-ph/0610274; 0610362.
 
 \bibitem{Zhu}S.H. Zhu, arXiv:hep-ph/0611270.


\bibitem{BJU_NPB} A.J. Buras, S. J$\rm \ddot{a}$ger and J. Urban, Nucl.
Phys. B {\bf 605}, 600 (2001).

\bibitem{LF} H.Y. Cheng {\it et al.}, Phys. Rev. D{\bf 69}, 074025
(2004).

\bibitem{KP} G. Kramer and W.F. Palmer, Phys. Rev. D{\bf 45}, 193
(1992); C.H. Chen and C.Q. Geng, Phys. Rev. D{\bf 71}, 115004
(2005); C.H. Chen and H. Htanaka, Phys. Rev. D{\bf 73}, 075003
(2006); C.H. Chen, arXiv:hep-ph/0601019.

\bibitem{Chen_1} C.H. Chen {\it et al.},  Phys. Rev. D{\bf 72}, 054011
(2005).

\bibitem{GLZ_PRD} C.Q. Geng {\it et al.}, Phys. Rev. D{\bf 62}, 074017
(2000).


\bibitem{HFAG} Heavy Flavor Averaging Group, E. Barberio {\it et
al.}, arXiv:hep-ex/0603003.


\bibitem{belle-tau} Belle Collaboration, K. Abe {\it et al.}, Phys. Rev. Lett. {\bf 92}, 171802
(2004); K. Hayasaka {\it et al.}, Phys. Lett. B{\bf 613}, 20 (2005).

\bibitem{babar-tau} Babar Collaboration, B. Aubert {\it et al.},
Phys. Rev. Lett. {\bf 95}, 041802 (2005); Phys. Rev. Lett. {\bf 96},
041801 (2006).

\bibitem{Suzuki} M. Suzuki, Phys. Rev. D{\bf 47}, 1252 (1993).

\bibitem{PDG_06}Particle Data Group, W.M. Yao {\it et al.}, J. Phys. G {\bf 33}, 1 (2006).

\bibitem{BW}P. Ball and R. Zwicky, Phys. Rev. D{\bf 71}, 014029
(2005).

\bibitem{D0} D0 Collaboration, V. Abazov {\it et al.}, Phys. Rev. D{\bf 74}, 031107
(2006).

\bibitem{GL} C.Q. Geng and C.C. Liu,
  J.\ Phys.\ G {\bf 29}, 1103 (2003).

\bibitem{GLW} C.Q. Geng, C.C. Lih and  W.W. Zhang, Phys. Rev. D{\bf 62}, 074017
(2000).


\end{thebibliography}
\end{document}